\documentclass[amsmath,amssymb,notitlepage,nofootinbib]{revtex4-1}
\usepackage{graphicx}
\usepackage{tikz-cd}
\usetikzlibrary{babel}
\usepackage{hyperref}
\usepackage{amsfonts}
\usepackage{amsthm}
\usepackage{mathtools}
\usepackage{amscd}
\usepackage{bm}
\usepackage{mathrsfs}
\usepackage[parfill]{parskip}    
\newcommand{\R}{\mathbb{R}}

\newcommand{\dR}{\mathrm{dR}}
\newcommand{\sing}{\mathrm{sing}}
\newcommand{\rel}{\mathrm{rel}}
\newcommand{\abs}{\mathrm{abs}}

\DeclareMathOperator{\im}{im}
\DeclareMathOperator{\Hom}{Hom}

\theoremstyle{definition}

\begin{document}
\title{Topological aspects of zero modes in cavity resonators}
\author{Osamu Kamigaito}
\email{kamigait@riken.jp}
\affiliation{RIKEN Nishina Center for Accelerator-Based Science\\
2-1 Hirosawa, Wako-shi, Saitama 351-0198, Japan}
\begin{abstract}
We discuss the relationship between the zero modes of electromagnetic fields in a cavity resonator and the cavity's topological characteristics.
We show that the dimension of the electromagnetic zero-mode space coincides with the dimension of the corresponding homology group of the cavity.
Moreover, we prove that the alternating sum of the dimensions of the electromagnetic zero-mode spaces is closely related to the Euler characteristic of the cavity boundary, 
and hence to the integral of the curvature.
\end{abstract}
%
\maketitle
\section{Introduction}

A cavity resonator is a region of space enclosed by metallic walls. Under suitable boundary conditions, electromagnetic fields can exist in the cavity only at discrete (resonant) frequencies.

To treat electromagnetic fields in a cavity in a mathematical way, we consider the eigenvalue problems for the operator
$\nabla\times\nabla\times-\nabla\nabla\cdot$
under the following two types of boundary conditions \cite{bib:kur69,bib:dom92}.
One set is
\begin{eqnarray}
\label{eq:e}
\nabla\times\nabla\times\bm{E}-\nabla\nabla\cdot\bm{E}
-k^2\bm{E}=0&\qquad \quad {\mbox {in V}}& \\
\label{eq:ebou}
\bm{n}\times\bm{E}=0,\quad
\nabla\cdot\bm{E}=0&\qquad \quad{\mbox {on S}},&
\end{eqnarray}
and the other set is
\begin{eqnarray}
\label{eq:h}
\nabla\times\nabla\times\bm{H}-\nabla\nabla\cdot\bm{H}
-k^2\bm{H}=0&\qquad \quad {\mbox {in V}} \\
\label{eq:hbou}
\bm{n}\cdot\bm{H}=0,\quad
\bm{n}\times\nabla\times\bm{H}=0&\qquad \quad{\mbox {on S}}.&
\end{eqnarray}
Here, $V$ is the interior of the cavity, $S$ is the cavity boundary, and $\bm{n}$ is the outward unit normal vector on $S$.
Among the solutions $\{\bm{E}_n\}$ and $\{\bm{H}_m\}$ obtained from the above equations, those that satisfy zero divergence in $V$ can be regarded as representing electromagnetic fields inside the cavity.
Moreover, the nonzero eigenvalues are common to the two eigenvalue problems, and the corresponding eigenvector fields are mapped to each other by $\nabla\times$.
A detailed explanation of this correspondence is given in the Appendix~\ref{sec:em}.

In contrast, the eigenmodes corresponding to the zero eigenvalue represent electrostatic and magnetostatic fields.
They can exist independently and do not form pairs.
It is also intuitively clear that the degeneracy of the zero modes is equal to the number of isolated conductors inside the region and to the number of ring-shaped current paths, respectively.
In this way, the zero modes reflect the topology of the cavity.

In this paper, we study the relationship between electromagnetic zero modes and the topology of a cavity.
We first introduce differential forms and treat the zero modes as harmonic fields. Using the Hodge decomposition for differential forms, together with cohomology and homology on manifolds with boundary, we extract topological and geometrical features of the electromagnetic zero modes.
In the Appendix, we summarize general facts about resonant modes in a cavity and their description using differential forms. We also list the differential forms and related formulas used in this paper.
The main results of this paper can be found in \cite{bib:gil84,bib:ber03}.
Our viewpoint on differential forms on manifolds with boundary follows \cite{bib:sch95}.
\section{Settings with differential forms}

To develop the discussion, we rewrite the equations and boundary conditions in the previous section, which were written in the notation of three-dimensional vector calculus, in terms of differential forms, which are a mathematically useful tool.

First, let the $1$-form corresponding to the vector field $\bm{E}$ in (\ref{eq:e}) and (\ref{eq:ebou}) be
$\omega^1_E\coloneqq \tilde{g}(\bm{E})$,
and let the $1$-form corresponding to the vector field $\bm{H}$ in (\ref{eq:h}) and (\ref{eq:hbou}) be
$\omega^1_H\coloneqq \tilde{g}(\bm{H})$.
Then (\ref{eq:e}) and (\ref{eq:ebou}) can be written, using $\Delta\coloneqq d\delta+\delta d$, as follows:
\begin{eqnarray}
\label{eq:lape}
\Delta\omega^1_E-k^2\omega^1_E=0&\qquad \quad {\mbox {in V}}& \\
\label{eq:boulape}
i^{\ast}\omega^1_E=0,\quad
i^{\ast}\delta\omega^1_E=0&\qquad \quad{\mbox {on S}}.&
\end{eqnarray}
Here $i^{\ast}: \Omega^p(M)\rightarrow \Omega^p(\partial M)$ is the pullback of the inclusion map $i:\partial M\rightarrow M$.
We call (\ref{eq:boulape}) the \emph{relative boundary condition}.
Similarly, (\ref{eq:h}) and (\ref{eq:hbou}) become
\begin{eqnarray}
\label{eq:laph}
\Delta\omega^1_H-k^2\omega^1_H=0&\qquad \quad {\mbox {in V}}& \\
\label{eq:boulaph}
i^{\ast}\star\omega^1_H=0,\quad
i^{\ast}\star d\omega^1_H=0&\qquad \quad{\mbox {on S}}.&
\end{eqnarray}
We call (\ref{eq:boulaph}) the \emph{absolute boundary condition}.

Equations (\ref{eq:laph}) and (\ref{eq:boulaph}) can be rewritten by introducing
$\omega^2_B\coloneqq \star\omega^1_H$ and using basic properties of the Hodge star operator. We obtain
\begin{eqnarray}
\Delta\omega^2_B-k^2\omega^2_B=0&\qquad \quad {\mbox {in V}}& \\
i^{\ast}\omega^2_B=0,\quad
i^{\ast}\delta\omega^2_B=0&\qquad \quad{\mbox {on S}}.&
\end{eqnarray}
Note that $\omega^2_B$ satisfies the relative boundary condition.
Similarly, equations (\ref{eq:lape}) and (\ref{eq:boulape}) can be rewritten by introducing
$\omega^2_D\coloneqq \star\omega^1_E$. Then we have
\begin{eqnarray}
\Delta\omega^2_D-k^2\omega^2_D=0&\qquad \quad {\mbox {in V}}& \\
i^{\ast}\star\omega^2_D=0,\quad
i^{\ast}\star d\omega^2_D=0&\qquad \quad{\mbox {on S}},&
\end{eqnarray}
and the eigenvalue problem is converted into the one with the absolute boundary condition.

If we extend the above discussion to general dimension, then for a $p$-form
$\omega\in \Omega^p(M)$ on an $n$-dimensional manifold $M$ with boundary, the eigenvalue problem for $\Delta$ under the relative boundary condition is written as
\begin{eqnarray}
\Delta\omega-k^2\omega=0&&\qquad \quad \mbox{on}~M \\
\label{eq:bourel}
i^{\ast}\omega=0,\quad
i^{\ast}\delta\omega=0&&\qquad \quad\mbox{on}~\partial M.
\end{eqnarray}
Similarly, the eigenvalue problem for $\Delta$ under the absolute boundary condition can be written as
\begin{eqnarray}
\Delta\omega-k^2\omega=0&&\qquad \quad \mbox{on}~M \\
\label{eq:bouabs}
i^{\ast}\star\omega=0,\quad
i^{\ast}\star d\omega=0&&\qquad \quad\mbox{on}~\partial M.
\end{eqnarray}

In the following sections, we focus on the case $k=0$, i.e., the zero modes.

\section{Harmonic forms and harmonic fields}

In what follows, we assume that the manifold $M$ with boundary is orientable and compact, and that all differential forms on $M$ are smooth and integrable.
First, for a $k$-form $\omega$ on $M$, we call the space of all forms satisfying
\begin{eqnarray}
\Delta\omega=0
\end{eqnarray}
the space of \emph{$k$-harmonic forms}, denoted by $\mathscr{H}^k$.
That is, we set
\begin{eqnarray}
\mathscr{H}^k(M)\coloneqq\{\omega\in \Omega^k(M)~|~\Delta\omega=0\}.
\end{eqnarray}
On the other hand, a differential form satisfying $d\omega=0$ and $\delta\omega=0$ is called a \emph{harmonic field}.
We denote the space of such forms by
\begin{eqnarray}
\label{eq:hfield}
\mathcal{H}^k(M)\coloneqq\{\omega\in \Omega^k(M)~|~d\omega=0,\delta\omega=0\}.
\end{eqnarray}

In general, we have $\mathcal{H}^k(M)\subset \mathscr{H}^k(M)$, but on a manifold with boundary the reverse inclusion does not hold in general.
This is because Green's formula gives
\begin{eqnarray}
(\Delta \omega, \omega)_M=(d\omega, d\omega)_M+(\delta\omega, \delta\omega)_M+\int_{\partial M} i^{\ast}[\delta\omega\wedge\star\omega-\omega\wedge\star d\omega],
\end{eqnarray}
and the boundary integral does not necessarily vanish.
In the following cases, however, the boundary integral vanishes, and we recover the reverse inclusion.
Namely, a harmonic form becomes a harmonic field when it satisfies either the relative boundary condition (\ref{eq:bourel}),
\begin{eqnarray}
i^{\ast}\omega=0,\qquad i^{\ast}\delta\omega=0,
\end{eqnarray}
or the absolute boundary condition (\ref{eq:bouabs}),
\begin{eqnarray}
i^{\ast}\star\omega=0,\qquad i^{\ast}\star d\omega=0.
\end{eqnarray}

Therefore, let us define the relative harmonic forms $\mathscr{H}^k_{\rel}(M)$, the absolute harmonic forms $\mathscr{H}^k_{\abs}(M)$, the Dirichlet harmonic fields $\mathcal{H}^k_D(M)$, and the Neumann harmonic fields $\mathcal{H}^k_N(M)$ by
\begin{eqnarray}
&&\mathscr{H}^k_{\rel}(M)\coloneqq\{\omega\in \Omega^k(M)~|~\Delta\omega=0, i^{\ast}\omega=0, i^{\ast}\delta\omega=0\}\\
&&\mathscr{H}^k_{\abs}(M)\coloneqq\{\omega\in \Omega^k(M)~|~\Delta\omega=0, i^{\ast}\star\omega=0, i^{\ast}\star d\omega=0\}\\
\label{eq:harmd}
&&\mathcal{H}^k_D(M)\coloneqq\{\omega\in \Omega^k(M)~|~d\omega=0, \delta\omega=0, i^{\ast}\omega=0\}\\
\label{eq:harmn}
&&\mathcal{H}^k_N(M)\coloneqq\{\omega\in \Omega^k(M)~|~d\omega=0, \delta\omega=0, i^{\ast}\star\omega=0\}.
\end{eqnarray}
Then the following equalities hold:
\begin{eqnarray}
\mathscr{H}^k_{\rel}(M)=\mathcal{H}^k_D(M)\\
\mathscr{H}^k_{\abs}(M)=\mathcal{H}^k_N(M).
\end{eqnarray}
Therefore, the relative harmonic forms $\omega^1_E$ and $\omega^2_B$ are Dirichlet harmonic fields.
Likewise, the absolute harmonic forms $\omega^1_H$ and $\omega^2_D$ are Neumann harmonic fields.
In particular, all of these satisfy $d\omega=0$ and $\delta\omega=0$.
In terms of the corresponding vector fields, this means that both the curl and the divergence vanish.
Thus, it is natural to identify these harmonic fields with electrostatic or magnetostatic fields.

\section{De Rham cohomology}

Differential forms reflect the geometric structure of the space on which they are defined.
A key tool for seeing this is de Rham cohomology.
For differential forms on a manifold with boundary, one can define four cochain complexes and the corresponding cohomologies \cite{bib:sch95}.
For the linear maps $\mathbf{t}$ and $\mathbf{n}$ used in this section and the next, see Appendix~\ref{subsec:boundary}.

Let us start with the usual de Rham complex $(\Omega^{\ast}(M), d)$.
Here $\Omega^k(M)$ denotes the space of all $k$-forms on $M$, and the complex is
\begin{equation}
0\rightarrow \Omega^0(M)
\xrightarrow{d} \Omega^1(M)
\xrightarrow{d} \cdots
\xrightarrow{d}  \Omega^n(M)
\xrightarrow{d} 0.
\end{equation}
We define the $k$-th cohomology group of this complex by
\begin{eqnarray}
\mathbf{H}^k(M,d)\coloneqq \mbox{ker}~d|_{\Omega^k(M)}/\mbox{im}~d|_{\Omega^{k-1}(M)}.
\end{eqnarray}
This is the usual de Rham cohomology group:
\begin{eqnarray}
\label{eq:abscohom}
H^k_{\dR}(M)\coloneqq\mathbf{H}^k(M,d).
\end{eqnarray}

Next, we consider the dual complex of the de Rham complex, $(\Omega^{\ast}(M), \delta)$, given by
\begin{equation}
0\xleftarrow{\delta} \Omega^0(M)
\xleftarrow{\delta} \Omega^1(M)
\xleftarrow{\delta} \cdots
\xleftarrow{\delta}  \Omega^n(M)
\leftarrow 0.
\end{equation}
We define its $k$-th cohomology group by
\begin{eqnarray}
\mathbf{H}^k(M,\delta)\coloneqq \mbox{ker}~\delta|_{\Omega^k(M)}/\mbox{im}~\delta|_{\Omega^{k+1}(M)}.
\end{eqnarray}

Since $d$ commutes with $i^{\ast}$, it preserves the space of differential forms satisfying the 
\emph{Dirichlet boundary condition} $i^{\ast}\omega=0$.
Therefore, we set
\begin{eqnarray}
\label{eq:omegad}
\Omega^k_D(M)\coloneqq\{\omega\in\Omega^k(M)~|~i^{\ast}\omega=0\},
\end{eqnarray}
and construct the subcomplex $(\Omega^{\ast}_D(M), d)$ of $(\Omega^{\ast}(M), d)$ as
\begin{equation}
0\rightarrow \Omega^0_D(M)
\xrightarrow{d} \Omega^1_D(M)
\xrightarrow{d} \cdots
\xrightarrow{d}  \Omega^n_D(M)
\xrightarrow{d} 0.
\end{equation}
We define its $k$-th cohomology group by
\begin{eqnarray}
\mathbf{H}^k_r(M)\coloneqq \mbox{ker}~d|_{\Omega^k_D(M)}/\mbox{im}~d|_{\Omega^{k-1}_D(M)}.
\end{eqnarray}
This is called the relative de Rham cohomology group:
\begin{eqnarray}
\label{eq:relcohom}
H^k_{\dR}(M, \partial M)\coloneqq \mathbf{H}^k_r(M).
\end{eqnarray}

On the other hand, since $\delta$ commutes with $\mathbf{n}$, it preserves the space of differential forms satisfying the \emph{Neumann boundary condition}
$i^{\ast}\star\omega=0 \iff \mathbf{n}\omega=0$.
Therefore, we set
\begin{eqnarray}
\label{eq:omegan}
\Omega^k_N(M)\coloneqq\{\omega\in\Omega^k(M)~|~i^{\ast}\star\omega=0\},
\end{eqnarray}
and we can construct the subcomplex $(\Omega^{\ast}_N(M), \delta)$ of $(\Omega^{\ast}(M), \delta)$ as
\begin{equation}
0\xleftarrow{\delta} \Omega^0_N(M)
\xleftarrow{\delta} \Omega^1_N(M)
\xleftarrow{\delta} \cdots
\xleftarrow{\delta}  \Omega^n_N(M)
\leftarrow 0.
\end{equation}
We define its $k$-th cohomology group by
\begin{eqnarray}
\mathbf{H}^k_a(M)\coloneqq \mbox{ker}~\delta|_{\Omega^k_N(M)}/\mbox{im}~\delta|_{\Omega^{k+1}_N(M)}.
\end{eqnarray}

\section{Hodge isomorphism}

To compute the de Rham cohomology groups introduced above, we use the Hodge decomposition on manifolds with boundary (the Hodge--Morrey--Friedrichs decomposition) \cite{bib:sch95}.
There are two different orthogonal decompositions, depending on how the harmonic fields are split:
\begin{eqnarray}
\label{eq:hmfn}
\Omega^k(M)=d\Omega^{k-1}_D(M)\oplus\delta\Omega^{k+1}_N(M)\oplus\mathcal{H}^k_{\mathrm{ex}}(M)\oplus\mathcal{H}^k_N(M)\\
\label{eq:hmfd}
\Omega^k(M)=d\Omega^{k-1}_D(M)\oplus\delta\Omega^{k+1}_N(M)\oplus\mathcal{H}^k_{\mathrm{co}}(M)\oplus\mathcal{H}^k_D(M).
\end{eqnarray}
Here $\Omega^k_D(M)$, $\Omega^k_N(M)$, $\mathcal{H}^k_D(M)$, and $\mathcal{H}^k_N(M)$ are defined in (\ref{eq:omegad}), (\ref{eq:omegan}), (\ref{eq:harmd}), and (\ref{eq:harmn}), respectively.
Also, letting $\mathcal{H}^k(M)$ be the space of harmonic fields in (\ref{eq:hfield}), we define
\begin{eqnarray}
\mathcal{H}^k_{\mathrm{ex}}(M)\coloneqq\{\kappa\in\mathcal{H}^k(M)~|~\kappa=d\epsilon\}\\
\mathcal{H}^k_{\mathrm{co}}(M)\coloneqq\{\kappa\in\mathcal{H}^k(M)~|~\kappa=\delta\gamma\}.
\end{eqnarray}

Using this decomposition, we obtain the following Hodge isomorphisms (\cite{bib:sch95}, Theorem 2.6.1):
\begin{eqnarray}
\label{eq:hodgen0}
\mathbf{H}^k(M, d)\cong\mathcal{H}^k_N(M)\\
\label{eq:hodged0}
\mathbf{H}^k(M,\delta)\cong\mathcal{H}^k_D(M).
\end{eqnarray}
Moreover,
\begin{eqnarray}
\label{eq:hodged}
\mathbf{H}^k_r(M)\cong\mathcal{H}^k_D(M)\\
\label{eq:hodgen}
\mathbf{H}^k_a(M)\cong\mathcal{H}^k_N(M)
\end{eqnarray}
also holds (\cite{bib:sch95}, pp.~103--104).
Below we verify (\ref{eq:hodged}) and (\ref{eq:hodgen}).
\begin{proof}
We first prove (\ref{eq:hodged}). By (\ref{eq:hmfd}), any $\omega\in\Omega^k(M)$ can be decomposed as
\begin{eqnarray}
\label{eq:eq1}
\omega=d\alpha_{\omega}+\delta\beta_{\omega}+\delta\gamma_{\omega}+\lambda_{\omega},
\end{eqnarray}
where $\alpha_{\omega}\in\Omega^{k-1}_D(M)$, $\beta_{\omega}\in\Omega^{k+1}_N(M)$,
$\delta\gamma_{\omega}\in\mathcal{H}^k_{\mathrm{co}}(M)$, and $\lambda_{\omega}\in\mathcal{H}^k_D(M)$.
If we assume $d\omega=0$, then we get $d\delta\beta_{\omega}=0$.
Applying Green's formula, we obtain
\begin{eqnarray*}
0=(d\delta\beta_{\omega}, \beta_{\omega})_M=(\delta\beta_{\omega}, \delta\beta_{\omega})_M
+\int_{\partial M} i^{\ast}(\delta\beta_{\omega}\wedge\star\beta_{\omega}).
\end{eqnarray*}
Since $i^{\ast}\star\beta_{\omega}=0$, the second term on the right-hand side is zero, and hence $\delta\beta_{\omega}=0$.
Therefore, the decomposition (\ref{eq:eq1}) reduces to $\omega=d\alpha_{\omega}+\delta\gamma_{\omega}+\lambda_{\omega}$.
If we further assume $\omega\in\Omega^k_D(M)$, then $\mathbf{t}\omega=\mathbf{t}\delta\gamma_{\omega}=0$, that is,
$i^{\ast}\delta\gamma_{\omega}=0$.
Since we already have $d\delta\gamma_{\omega}=0$, Green's formula gives
\begin{eqnarray*}
0=(d\delta\gamma_{\omega}, \gamma_{\omega})_M=(\delta\gamma_{\omega}, \delta\gamma_{\omega})_M
+\int_{\partial M} i^{\ast}(\delta\gamma_{\omega}\wedge\star\gamma_{\omega}).
\end{eqnarray*}
By what we just showed, the second term on the right-hand side is zero, so we conclude that $\delta\gamma_{\omega}=0$.
From these facts, when $\omega\in\ker d|_{\Omega^k_D(M)}$, we can write $\omega=d\alpha_{\omega}+\lambda_{\omega}$.
This proves (\ref{eq:hodged}).
Next we prove (\ref{eq:hodgen}). By (\ref{eq:hmfn}), any $\omega\in\Omega^k(M)$ can be decomposed as
\begin{eqnarray}
\label{eq:eq2}
\omega=d\alpha_{\omega}+\delta\beta_{\omega}+d\epsilon_{\omega}+\kappa_{\omega},
\end{eqnarray}
where $\alpha_{\omega}\in\Omega^{k-1}_D(M)$, $\beta_{\omega}\in\Omega^{k+1}_N(M)$,
$d\epsilon_{\omega}\in\mathcal{H}^k_{\mathrm{ex}}(M)$, and $\kappa_{\omega}\in\mathcal{H}^k_N(M)$.
If we assume $\delta\omega=0$, then we get $\delta d\alpha_{\omega}=0$.
Applying Green's formula, we obtain
\begin{eqnarray*}
0=( \alpha_{\omega}, \delta d\alpha_{\omega})_M=(d\alpha_{\omega}, d\alpha_{\omega})_M
-\int_{\partial M} i^{\ast}(\alpha_{\omega}\wedge\star d\alpha_{\omega}).
\end{eqnarray*}
Since $i^{\ast}\alpha_{\omega}=0$, the second term on the right-hand side is zero, and hence $d\alpha_{\omega}=0$.
Therefore, the decomposition (\ref{eq:eq2}) reduces to $\omega=\delta\beta_{\omega}+d\epsilon_{\omega}+\kappa_{\omega}$.
If we further assume $\omega\in\Omega^k_N(M)$, then $\mathbf{n}\omega=\mathbf{n}d\epsilon_{\omega}=0$, that is,
$i^{\ast}\star d\epsilon_{\omega}=0$.
Since we already have $\delta d\epsilon_{\omega}=0$, Green's formula gives
\begin{eqnarray*}
0=(\epsilon_{\omega}, \delta d\epsilon_{\omega})_M=(d\epsilon_{\omega}, d\epsilon_{\omega})_M
-\int_{\partial M} i^{\ast}(\epsilon_{\omega}\wedge\star d\epsilon_{\omega}).
\end{eqnarray*}
By what we just showed, the second term on the right-hand side is zero, so we conclude that $d\epsilon_{\omega}=0$.
From these facts, when $\omega\in\ker \delta|_{\Omega^k_N(M)}$, we can write $\omega=\delta\beta_{\omega}+\kappa_{\omega}$.
This proves (\ref{eq:hodgen}).
\end{proof}

Combining (\ref{eq:relcohom}) and (\ref{eq:hodged}), we obtain
\begin{eqnarray}
\label{eq:hodgeisod}
H^k_{\dR}(M, \partial M)\cong\mathcal{H}^k_D(M).
\end{eqnarray}
Also, combining (\ref{eq:abscohom}) and (\ref{eq:hodgen0}), we obtain
\begin{eqnarray}
\label{eq:hodgeison}
H^k_{\dR}(M)\cong\mathcal{H}^k_N(M).
\end{eqnarray}
Moreover, the Hodge star operator $\star:\mathcal{H}^k_N(M)\rightarrow \mathcal{H}^{n-k}_D(M)$ induces an isomorphism
$\star_{\mathrm{P}}:H^k_{\dR}(M)\rightarrow H^{n-k}_{\dR}(M, \partial M)$:

\begin{equation}
\label{eq:pl}
\begin{tikzcd}
  \mathcal{H}^k_N(M) \ar[r, "\star"] \arrow[d, "\cong"'] 
  & \mathcal{H}^{n-k}_D(M) \ar[d, "\cong"] \\
  H^k_{\dR}(M) \ar[r, "\star_{\mathrm{P}}"] 
  & H^{n-k}_{\dR}(M, \partial M)
\end{tikzcd}
\end{equation}

Since $\mathcal{H}^k_D(M)$ is finite-dimensional (\cite{bib:sch95}, Theorem 2.2.2), (\ref{eq:hodgeisod}) implies that $H^k_{\dR}(M, \partial M)$ is also finite-dimensional.
Moreover, (\ref{eq:pl}) shows that $\mathcal{H}^k_N(M)$ is finite-dimensional as well, and hence (\ref{eq:hodgeison}) implies that $H^k_{\dR}(M)$ is finite-dimensional.

\section{Poincar\'{e} duality on manifolds with boundary}

Using the Hodge isomorphisms in the previous section, we describe the duality that holds between de Rham cohomology groups (\cite{bib:mor01} \S 4.4(a)).
For $\omega\in\ker d|_{\Omega^k_D(M)}$ and $\eta\in\ker d|_{\Omega^{n-k}(M)}$ on an $n$-dimensional manifold $M$ with boundary, we can define a bilinear form
\[
\widehat{B}:\ker d|_{\Omega^k_D(M)}\times \ker d|_{\Omega^{n-k}(M)}\longrightarrow \R,
\qquad
\widehat{B}(\omega, \eta) \coloneqq \int_M \omega\wedge \eta.
\]
If we take $\alpha\in\Omega^{k-1}_D(M)$ and $\beta\in\Omega^{n-k-1}(M)$, then
\begin{eqnarray}
\widehat{B}(\omega+d\alpha, \eta+d\beta)&=& \int_M (\omega+d\alpha)\wedge (\eta+d\beta)\nonumber\\
&=& \int_M \omega\wedge\eta+ \int_M d(\alpha\wedge\eta+(-1)^k\omega\wedge\beta+\alpha\wedge d\beta)\nonumber\\
&=& \int_M \omega\wedge\eta+ \int_{\partial M} i^{\ast}(\alpha\wedge\eta+(-1)^k\omega\wedge\beta+\alpha\wedge d\beta).
\end{eqnarray}
Since $i^{\ast}\alpha=0$ and $i^{\ast}\omega=0$, the second term is zero.
Therefore,
\begin{eqnarray}
\widehat{B}(\omega+d\alpha, \eta+d\beta)=\widehat{B}(\omega, \eta),
\end{eqnarray}
and hence we obtain an induced pairing
\[
B: H^k_{\dR}(M,\partial M)\times H^{n-k}_{\dR}(M)\longrightarrow \R,
\qquad
B([\omega],[\eta])=\widehat{B}(\omega, \eta)=\int_M \omega\wedge \eta.
\]

This bilinear form $B$ is nondegenerate, and it induces the following isomorphism:
\begin{eqnarray}
\label{eq:isohodge}
H^k_{\dR}(M, \partial M)\cong \bigl(H^{n-k}_{\dR}(M)\bigr)^{\ast}.
\end{eqnarray}
\begin{proof}
Take a nonzero class $[\omega]\in H^k_{\dR}(M,\partial M)$.
By the Hodge isomorphism (\ref{eq:hodgeisod}), we may choose a representative $\omega\in\mathcal{H}^k_D(M)$.
On the other hand, using the isomorphism $\star_{\mathrm{P}}$ defined in the diagram (\ref{eq:pl}) and the Hodge star operator $\star$, we have
$\star_{\mathrm{P}}[\omega]=[\star\omega]\in H^{n-k}_{\dR}(M)$, and
$\star\omega\in\mathcal{H}^{n-k}_N(M)$.
Then
\begin{eqnarray}
B([\omega],\star_{\mathrm{P}}[\omega])=B([\omega],[\star\omega])=\widehat{B}(\omega,\star\omega)
=\int_M \omega\wedge \star\omega=(\omega,\omega)_M>0.
\end{eqnarray}
Therefore, $B$ is nondegenerate.
Define $\widetilde{B}:H^k_{\dR}(M, \partial M)\rightarrow \bigl(H^{n-k}_{\dR}(M)\bigr)^{\ast}$ by
\begin{eqnarray}
\widetilde{B}([\omega])([\eta])\coloneqq B([\omega],[\eta]) \quad\mathrm{for}~[\eta]\in H^{n-k}_{\dR}(M).
\end{eqnarray}
Then $\widetilde{B}$ is an isomorphism.
\end{proof}

\section{De Rham's theorem}

Next we outline the de Rham theorem (\cite{bib:mor01} \S 3.3(b)).
This will make it clear that there is an isomorphism between de Rham cohomology and singular cohomology.
Let $M$ be an $n$-dimensional manifold with boundary.
A smooth map $\sigma:\Delta^k\rightarrow M$ from the standard $k$-simplex $\Delta^k$ to $M$ is called a singular $k$-simplex.
Let $C_k^{\sing}(M)$ be the free abelian group over $\R$ generated by all singular $k$-simplices, and call its elements singular $k$-chains in $M$.
With the boundary operator denoted by $\partial$, we have $\partial c\in C_{k-1}^{\sing}(M)$ for any $c\in C_k^{\sing}(M)$.
The collection $\{C_k^{\sing}(M), \partial\}$, that is,
\begin{equation}
0\rightarrow C_n^{\sing}(M)
\xrightarrow{\partial} C_{n-1}^{\sing}(M)
\xrightarrow{\partial} \cdots
\xrightarrow{\partial} C_1^{\sing}(M)
\xrightarrow{\partial} C_0^{\sing}(M)
\rightarrow 0,
\end{equation}
is called the singular chain complex of $M$.
Define
$Z_k^{\sing}(M)\coloneqq\{c\in C_k^{\sing}(M);\ \partial c=0\}$ and
$B_k^{\sing}(M)\coloneqq\{\partial c\in C_{k+1}^{\sing}(M)\}$.
Then $B_k^{\sing}(M)\subset Z_k^{\sing}(M)$.
The quotient
$H_k^{\sing}(M)\coloneqq Z_k^{\sing}(M)/B_k^{\sing}(M)$
is called the $k$-th singular homology group of $M$.

We next introduce the singular cochain complex $\{C^k_{\sing}(M), \delta\}$, which is dual to the singular chain complex $\{C_k^{\sing}(M), \partial\}$, by
$C^k_{\sing}(M)\coloneqq \Hom(C_k^{\sing}(M), \R)$.
Here $\delta$ is the coboundary operator dual to the boundary map, defined by
\[
\delta f(c)\coloneqq f(\partial c)
\qquad
\text{for } f\in C^k_{\sing}(M),\ c\in C_{k+1}^{\sing}(M).
\]
The identity $\delta\cdot\delta=0$ follows from $\partial\cdot\partial=0$:
\begin{equation}
0\rightarrow C^0_{\sing}(M)
\xrightarrow{\delta} C^1_{\sing}(M)
\xrightarrow{\delta} \cdots
\xrightarrow{\delta} C^{n-1}_{\sing}(M)
\xrightarrow{\delta} C^n_{\sing}(M)
\rightarrow 0.
\end{equation}
Now define
$Z^k_{\sing}(M)\coloneqq\{f\in C^k_{\sing}(M);\ \delta f=0\}$ and
$B^k_{\sing}(M)\coloneqq\{\delta f\in C^{k-1}_{\sing}(M)\}$.
Then $B^k_{\sing}(M)\subset Z^k_{\sing}(M)$.
The quotient
$H^k_{\sing}(M)\coloneqq Z^k_{\sing}(M)/B^k_{\sing}(M)$
is called the $k$-th singular cohomology group of $M$.

For any $\omega\in\Omega^k(M)$, we define a singular $k$-cochain $I(\omega)\in C^k_{\sing}(M)$ as follows:
\begin{eqnarray}
I(\omega)(\sigma):=\int_{\Delta^k}\sigma^{\ast}\omega,
\end{eqnarray}
where $\sigma:\Delta^k\to M$ is a singular simplex.
The map $I:\Omega^k(M)\to C^k_{\sing}(M)$ satisfies
\begin{eqnarray}
\delta\, I(\omega)= I(d\omega),
\end{eqnarray}
where $\delta$ is the singular coboundary operator.
This means that a closed form and an exact form are mapped to a cocycle and a coboundary, respectively.
Therefore, $I$ is a cochain map, and it induces a map on cohomology,
\begin{eqnarray}
I^{\ast}:H^k_{\dR}(M)\to H^k_{\sing}(M).
\end{eqnarray}
The de Rham theorem states that $I^{\ast}$ is an isomorphism; in other words,
\begin{eqnarray}
\label{eq:drabs}
H^k_{\dR}(M)\cong H^k_{\sing}(M)
\end{eqnarray}
holds.

On the other hand, for a manifold with boundary, we consider the relative singular chain complex built from
$C_n^{\sing}(M)$, $C_n^{\sing}(\partial M)$, and
$C_n^{\sing}(M, \partial M)\coloneqq C_n^{\sing}(M)/C_n^{\sing}(\partial M)$.
For each $n$, these groups fit into the following short exact sequence:
\begin{equation}
0\rightarrow C_n^{\sing}(\partial M)
\xrightarrow{i} C_n^{\sing}(M)
\xrightarrow{\pi}  C_n^{\sing}(M, \partial M) 
\rightarrow 0.
\end{equation}
Here $i: C_n^{\sing}(\partial M)\rightarrow C_n^{\sing}(M)$ is the inclusion map and is injective, and
$\pi: C_n^{\sing}(M)\rightarrow C_n^{\sing}(M, \partial M)$ is the projection map and is surjective.
This short exact sequence induces the long exact sequence in relative homology \cite{bib:hat02}:
\begin{equation}
\label{eq:homlong}
0\rightarrow \cdots
\rightarrow H_k^{\sing}(\partial M)
\rightarrow H_k^{\sing}(M)
\rightarrow  H_k^{\sing}(M,\partial M)
\rightarrow H_{k-1}^{\sing}(\partial M)
\rightarrow \cdots
\rightarrow 0.
\end{equation}
We will discuss the geometric meaning of $H_k^{\sing}(M,\partial M)$ later.

For the relative chain complex, we now consider its dual, the relative singular cochain complex.
First, for each $k$, we introduce the pullback of the inclusion map $i$,
\begin{equation}
i^{\ast}:C^k_{\sing}(M)\to C^k_{\sing}(\partial M).
\end{equation}
We define the relative singular cochain complex by
\begin{equation}
C^k_{\sing}(M,\partial M):=
\{\varphi\in C^k_{\sing}(M)\mid \varphi|_{C_k(\partial M)}=0\}.
\end{equation}
In other words, a relative singular cochain is a cochain that vanishes on singular chains supported in $\partial M$.
Then for each $k$ there is a short exact sequence
\begin{equation}
\label{eq:ses-cochains}
0 \rightarrow  C^k_{\sing}(M,\partial M)
\rightarrow  C^k_{\sing}(M)
\xrightarrow{j^{\ast}} C^k_{\sing}(\partial M)
\rightarrow  0,
\end{equation}
which induces the long exact sequence
\begin{equation}
\label{eq:les-cochains}
0\rightarrow \cdots 
\to H^{k-1}_{\sing}(\partial M)
\xrightarrow{\delta} H^k_{\sing}(M,\partial M) 
\to H^k_{\sing}(M)
\to H^k_{\sing}(\partial M)
\xrightarrow{\delta} \cdots
\rightarrow 0.
\end{equation}
Now, if $\omega\in\Omega^k_D(M)$, that is, $i^{\ast}\omega=0$, then for any singular simplex
$\sigma:\Delta^k\to \partial M$ we have
\begin{equation}
I(\omega)(\sigma)=\int_{\Delta^k}\sigma^{\ast}\omega
=\int_{\Delta^k}\sigma^{\ast}(i^{\ast}\omega)=0.
\end{equation}
Therefore, $I(\omega)$ vanishes on simplices in $\partial M$, and the map
\begin{equation}
I:\Omega^k_D(M)\to C^k_{\sing}(M,\partial M)
\end{equation}
is well defined.

Moreover, since $I$ is a cochain map, that is, $\delta I = I d$, it induces a homomorphism on cohomology:
\begin{equation}
I_*:H^k_{\dR}(M,\partial M)\longrightarrow H^k_{\sing}(M,\partial M).
\end{equation}
The de Rham theorem states that this map is an isomorphism.
In other words, for a manifold with boundary $M$,
\begin{equation}
\label{eq:drrel}
H^k_{\dR}(M,\partial M)\cong H^k_{\sing}(M,\partial M)
\end{equation}
holds.

\section{Relationship with singular homology}

In the previous sections, we confirmed the isomorphisms between de Rham cohomology and singular cohomology over $\R$.
In this section, we study the relation between singular cohomology and singular homology over $\R$.
The key tool is Poincar\'e--Lefschetz duality (\cite{bib:hat02}, Theorem 3.43), which states that
\begin{eqnarray}
\label{eq:plabs}
&&H^k_{\sing}(M)\cong H_{n-k}^{\sing}(M,\partial M),\\
\label{eq:plrel}
&&H^k_{\sing}(M,\partial M)\cong H_{n-k}^{\sing}(M).
\end{eqnarray}
Using this, we can derive the following important relation.
For the left-hand side of (\ref{eq:isohodge}), by (\ref{eq:drrel}) and (\ref{eq:plrel}) we have
\begin{eqnarray}
H^k_{\dR}(M, \partial M)\cong H^k_{\sing}(M, \partial M)\cong H_{n-k}^{\sing}(M),
\end{eqnarray}
and for the right-hand side, by (\ref{eq:drabs}) and (\ref{eq:plabs}) we have
\begin{eqnarray}
\bigl(H^{n-k}_{\dR}(M)\bigr)^{\ast}\cong \bigl(H^{n-k}_{\sing}(M)\bigr)^{\ast}\cong \bigl(H_k^{\sing}(M, \partial M)\bigr)^{\ast}.
\end{eqnarray}
Therefore,
\begin{eqnarray}
H_{n-k}^{\sing}(M)\cong \bigl(H_k^{\sing}(M, \partial M)\bigr)^{\ast}.
\end{eqnarray}
Note that all homology and cohomology groups appearing here are finite-dimensional, by (\ref{eq:hodgeisod}), (\ref{eq:hodgeison}), and (\ref{eq:pl}).
In general, for a finite-dimensional vector space $V$, the dimension of its dual space $V^{\ast}$ equals the dimension of $V$: $\dim V=\dim V^{\ast}$.
Taking dimensions on both sides of the above isomorphism, we obtain
\begin{eqnarray}
\label{eq:resultpl}
\dim H_{n-k}^{\sing}(M)=\dim H_k^{\sing}(M, \partial M).
\end{eqnarray}

Another important relation follows as a corollary of the universal coefficient theorem (\cite{bib:kaw22}, Corollary 8.1.12):
\begin{eqnarray}
\label{eq:uctabs}
&&H^k_{\sing}(M)\cong \Hom\bigl(H_k^{\sing}(M),\R\bigr)=\bigl(H_k^{\sing}(M)\bigr)^{\ast}\\
\label{eq:uctrel}
&&H^k_{\sing}(M,\partial M)\cong \Hom\bigl(H_k^{\sing}(M,\partial M),\R\bigr)=\bigl(H_k^{\sing}(M,\partial M)\bigr)^{\ast}.
\end{eqnarray}
From (\ref{eq:hodgeisod}), (\ref{eq:drrel}), and (\ref{eq:uctrel}), we obtain the following isomorphisms:
\begin{eqnarray}
\label{eq:iso1}
\mathcal{H}^k_D(M)\cong H^k_{\sing}(M, \partial M)\cong \bigl(H_k^{\sing}(M,\partial M)\bigr)^{\ast}.
\end{eqnarray}
Here $\mathcal{H}^1_D(M)$ and $\mathcal{H}^2_D(M)$ correspond to the electrostatic field $\bm{E}$ and the magnetic flux density $\bm{B}$, respectively. That is,
\begin{eqnarray}
\label{eq:etorel}
\{\bm{E}\}&\Rightarrow& \mathcal{H}^1_D(M)\cong \bigl(H_1^{\sing}(M,\partial M)\bigr)^{\ast}\\
\label{eq:btorel}
\{\bm{B}\}&\Rightarrow& \mathcal{H}^2_D(M)\cong \bigl(H_2^{\sing}(M,\partial M)\bigr)^{\ast}.
\end{eqnarray}
Similarly, from (\ref{eq:hodgeison}), (\ref{eq:drabs}), and (\ref{eq:uctabs}), we obtain
\begin{eqnarray}
\label{eq:iso2}
\mathcal{H}^k_N(M)\cong H^k_{\sing}(M)\cong H_{n-k}^{\sing}(M, \partial M).
\end{eqnarray}
Moreover, by (\ref{eq:hodged}), (\ref{eq:drrel}), and (\ref{eq:uctabs}), we have
\begin{equation}
\mathcal{H}^k_N(M)\cong H^k_{\dR}(M)\cong H^k_{\sing}(M)\cong  \bigl(H_k^{\sing}(M)\bigr)^{\ast}.
\end{equation}
Here $\mathcal{H}^1_N(M)$ and $\mathcal{H}^2_N(M)$ correspond to the magnetostatic field $\bm{H}$ and the electric flux density $\bm{D}$, respectively. That is,
\begin{eqnarray}
\label{eq:htobas}
\{\bm{H}\}&\Rightarrow& \mathcal{H}^1_N(M)\cong \bigl(H_1^{\sing}(M)\bigr)^{\ast}\\
\label{eq:dtobas}
\{\bm{D}\}&\Rightarrow& \mathcal{H}^2_N(M)\cong \bigl(H_2^{\sing}(M)\bigr)^{\ast}.
\end{eqnarray}
Later we will relate these groups to geometric objects.

\section{Euler characteristics}

Here we define the Betti numbers of singular homology groups:
\begin{eqnarray}
&&b_k(M, \partial M)\coloneqq\dim H_k^{\sing}(M, \partial M)\\
&&b_k(M)\coloneqq\dim H_k^{\sing}(M)\\
&&b_k(\partial M)\coloneqq\dim H_k^{\sing}(\partial M).
\end{eqnarray}
From (\ref{eq:homlong}), we obtain the following relation for Euler characteristics:
\begin{eqnarray}
\chi(M, \partial M)=\chi(M)-\chi(\partial M).
\end{eqnarray}
Here
\begin{eqnarray}
&&\chi(M, \partial M)\coloneqq \sum_{k=0}^n (-1)^k b_k(M, \partial M)\\
&&\chi(M)\coloneqq \sum_{k=0}^n (-1)^k b_k(M)\\
&&\chi(\partial M)\coloneqq \sum_{k=0}^n (-1)^k b_k(\partial M).
\end{eqnarray}
\begin{proof}
Assume that the sequence
\[
0\rightarrow A^0
\xrightarrow{d_0} A^1
\xrightarrow{d_1}  A^2
\xrightarrow{d_2}  \cdots
\rightarrow A^m
\rightarrow 0
\]
is exact.
Then by a basic result in linear algebra,
\begin{eqnarray}
\dim A^k=\dim\ker d_k+\dim\im d_k
\end{eqnarray}
holds.
Taking the alternating sum and using $\dim\ker d_k=\dim\im d_{k-1}$, we obtain
\begin{eqnarray}
\sum_{k=0}^m (-1)^k \dim A^k=0.
\end{eqnarray}
Applying this to the long exact sequence in homology gives the desired relation.
\end{proof}

On the other hand, from (\ref{eq:resultpl}) we have
\begin{eqnarray}
b_k(M, \partial M)=b_{n-k}(M).
\end{eqnarray}
Therefore,
\begin{eqnarray}
\chi(M, \partial M)=(-1)^n\chi(M).
\end{eqnarray}
Substituting this into the previous relation, we obtain
\begin{eqnarray}
\chi(\partial M)=(1-(-1)^n)\chi(M).
\end{eqnarray}
In particular, when $n=3$ we have
\begin{eqnarray}
\chi(M, \partial M)=-\chi(M)=-\frac{1}{2}\chi(\partial M).
\end{eqnarray}
Since $\partial M$ is a closed $2$-dimensional surface, the Gauss curvature $K$ gives
\begin{eqnarray}
\chi(\partial M)=\frac{1}{2\pi}\int_{\partial M}K\, dA,
\end{eqnarray}
and finally we obtain
\begin{eqnarray}
\label{eq:final}
\chi(M, \partial M)=-\chi(M)=-\frac{1}{4\pi}\int_{\partial M}K\, dA.
\end{eqnarray}
It is interesting that the topological information of the electromagnetic fields inside the cavity is determined by the boundary (\cite{bib:gil84}, Theorem 4.2.7).

\section{Discussions}

\begin{figure}[htbp]
   \centering
   \includegraphics[width=14cm]{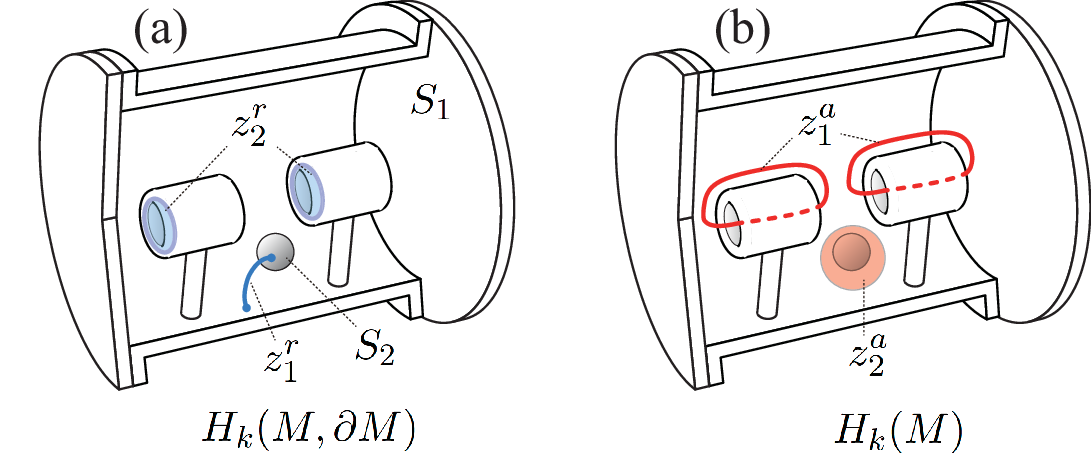} 
   \caption{Example of a resonant cavity. A cylindrical outer shell contains two drift tubes, and a metal sphere is floating in the vacuum.
$S_1$ is the inner surface of the cavity body, and $S_2$ is the surface of the sphere.
In (a) we show cycles $z_k^r$ representing classes in $H_k(M,\partial M)$, and in (b) we show cycles $z_k^a$ representing classes in $H_k(M)$.}
   \label{fig:cavity}
\end{figure}

An example of a resonant cavity is shown in Fig.~\ref{fig:cavity}.
Inside a cylindrical outer shell, two drift tubes are installed.
In addition, a metal sphere is floating in the vacuum (this is not realistic in practice, but a similar situation can be realized if the sphere is supported by a dielectric material).
Let $M$ be the interior of the cavity including its boundary, and let $\partial M$ be the inner surface of the cavity.
Assume that $\partial M$ consists of $N$ closed surfaces $S_1,\cdots, S_N$, so that
$\partial M=\sum_{j=1}^N S_j$.
Let $g_j$ be the genus of the closed surface $S_j$.
In the example in Fig.~\ref{fig:cavity}, we have $N=2$:
$S_1$ is the inner surface of the cavity body, $S_2$ is the surface of the sphere, and
$g_1=2$, $g_2=0$.

First, from (\ref{eq:etorel}), the group paired with $\{\bm{E}\}$ is $H_1(M,\partial M)$.
The curve $z_1^r$ shown in Fig.~\ref{fig:cavity}(a), which connects the sphere and the cavity wall, represents a class in $H_1(M,\partial M)$ because its boundary lies in $\partial M$.
As this example suggests, the relative Betti number, which gives the dimension of $\{\bm{E}\}$, is
$b_1(M,\partial M)=N-1$.
Next, the group paired with $\{\bm{B}\}$ is $H_2(M,\partial M)$.
The membrane $z_2^r$ shown in Fig.~\ref{fig:cavity}(a), which caps the hole of a drift tube, represents a class in $H_2(M,\partial M)$ because its boundary lies in $\partial M$.
As this example suggests, the relative Betti number, which gives the dimension of $\{\bm{B}\}$, is
$b_2(M,\partial M)=\sum_{j=1}^N g_j$.
Furthermore, in relative homology, the manifold $M$ itself is a cycle, and there is no $3$-boundary. Therefore the third relative homology group is
$H_3(M,\partial M)=\mathbb{R}$.
Also, in relative homology, any point can be realized as the boundary of a line segment whose other endpoint lies on the boundary. Therefore $H_0(M,\partial M)=0$.

Figure~\ref{fig:cavity}(b) shows the relation between $\{\bm{H}\}$, $\{\bm{D}\}$, and homology, in this case the ordinary (absolute) homology.
In particular, since $M$ is connected we have $H_0(M)=\mathbb{R}$, and since there are no $3$-cycles and no $3$-boundaries, the third homology group is trivial: $H_3(M)=0$.

\begin{table}[htbp]
\begin{center}
\caption{Summary of the Betti numbers of resonant cavities.}
\label{tab:betti}
\begin{tabular}{c|c|c}
$k$ &  $b_k(M,\partial M)$ &  $b_k(M)$\\
\hline
0  &0  & 1 \\
1 &  $N-1~(=\dim\{\bm{E}\})$ & $\sum_{j=1}^N g_j~(=\dim\{\bm{H}\})$\\
2 &  $\sum_{j=1}^N g_j~(=\dim\{\bm{B}\})$ & $N-1~(=\dim\{\bm{D}\})$\\
3  &1  & 0 \\
\end{tabular}
\end{center}
\end{table}

The corresponding Betti numbers are summarized in Table~\ref{tab:betti} \cite{bib:ber03}.
As shown in the table, we have
\begin{eqnarray}
\chi(M,\partial M)=-\chi(M)=\sum_{j=1}^N(g_j-1)=-\frac{1}{2}\sum_{j=1}^N\chi(S_j),
\end{eqnarray}
which agrees with (\ref{eq:final}).

\section*{Acknowledgements}
I gained substantial insight into viewing electromagnetism through differential forms from \cite{bib:yos19}.
I thank ChatGPT for help with English editing and for suggestions that improved the exposition.
This brief note ultimately traces back to \cite{bib:wey13}, which Dr.~Yoshiaki Chiba of the RILAC laboratory shared with me in 1993, shortly after I joined RIKEN.
Chiba-san told me then that the eigenvalue distribution of resonators is of fundamental importance.
I dedicate this note to his memory.

\appendix
\section{Electromagnetic fields in cavity resonators}
\label{sec:em}
In accelerator physics, an accelerating cavity can be regarded as a region enclosed by metallic walls (a closed surface) if we close the beam ports and power couplers.
The Maxwell equations in the cavity interior $V$ and the boundary conditions on the wall surface $S$ can be written as follows:
\begin{eqnarray}
\nabla\times\bm{E}+\mu_0\frac{\partial\bm{H}}{\partial t}=0 \qquad &&{\mbox {in V}}\\
\nabla\times\bm{H}-\epsilon_0\frac{\partial\bm{E}}{\partial t}=0 \qquad &&{\mbox {in V}}\\
\epsilon_0\nabla\cdot\bm{E}=0\qquad &&{\mbox {in V}}\\
\mu_0\nabla\cdot\bm{H}=0\qquad &&{\mbox {in V}}\\
\bm{n}\times\bm{E}=0\qquad &&{\mbox {on S}}\\
\bm{n}\cdot\bm{H}=0\qquad &&{\mbox {on S}}.
\end{eqnarray}
Here $\bm{n}$ is the outward unit normal vector on $S$.
Under these conditions, electromagnetic fields with discrete frequencies (eigenmodes) are allowed to exist in $V$.
The equations satisfied by such fields can be written in the following form, where the eigen angular frequency is $ck$ ($c$ is the speed of light):
\begin{eqnarray}
\nabla\times\nabla\times\bm{E}-k^2\bm{E}=0\qquad &&{\mbox {in V}} \\
\bm{n}\times\bm{E}=0\qquad &&{\mbox {on S}}\\
\nabla\times\nabla\times\bm{H}-k^2\bm{H}=0\qquad &&{\mbox {in V}} \\
\bm{n}\cdot\bm{H}=0\qquad &&{\mbox {on S}}.
\end{eqnarray}
Note that the equations for the electric field and the magnetic field have been separated.

To treat electromagnetic fields in a cavity in a mathematical way, it is convenient to consider, instead of the equations above, the eigenvalue problems for the vector Laplacian
$\nabla\times\nabla\times-\nabla\nabla\cdot$ \cite{bib:kur69,bib:dom92}.
The first set is
\begin{eqnarray}
\label{eq:efun1}
\nabla\times\nabla\times\bm{E}-\nabla\nabla\cdot\bm{E}
-k^2\bm{E}=0\qquad &&{\mbox {in V}} \\
\label{eq:efun2}
\bm{n}\times\bm{E}=0,\quad
\nabla\cdot\bm{E}=0\qquad &&{\mbox {on S}}.
\end{eqnarray}
Note that additional boundary conditions are imposed.
These eigenvector fields can be classified into the following three types, according to their properties in $V$:
\begin{eqnarray}
{\mbox {(A)}}\quad  \nabla\times\bm{E}_n=0,\qquad \quad \nabla\cdot\bm{E}_n=0\\
{\mbox {(B)}}\quad \nabla\times\bm{E}_n\neq0,\qquad \quad \nabla\cdot\bm{E}_n=0\\
{\mbox {(C)}}\quad \nabla\times\bm{E}_n=0,\qquad \quad \nabla\cdot\bm{E}_n\neq0
\end{eqnarray}
Here the subscript $n$ labels the eigenvector fields (eigenmodes).
The other set is
\begin{eqnarray}
\label{eq:bfun1}
\nabla\times\nabla\times\bm{H}-\nabla\nabla\cdot\bm{H}
-k^2\bm{H}=0\qquad &&{\mbox {in V}} \\
\label{eq:bfun2}
\bm{n}\cdot\bm{H}=0,\quad
\bm{n}\times\nabla\times\bm{H}=0\qquad &&{\mbox {on S}}
\end{eqnarray}
In the same way, these eigenvector fields can also be classified into the following three types:
\begin{eqnarray}
{\mbox {(A')}}\quad  \nabla\times\bm{H}_m=0,\qquad \quad \nabla\cdot\bm{H}_m=0\\
{\mbox {(B')}}\quad \nabla\times\bm{H}_m\neq0,\qquad \quad \nabla\cdot\bm{H}_m=0\\
{\mbox {(C')}}\quad \nabla\times\bm{H}_m=0,\qquad \quad \nabla\cdot\bm{H}_m\neq0
\end{eqnarray}
Among the fields $\{\bm{E}_n\}$ and $\{\bm{H}_m\}$ obtained above, the modes of types (B) and (B') have the following property:
the nonzero eigenvalues are common, and the corresponding eigenvector fields are mapped to each other by $\nabla\times$.
More precisely, take an eigenvector field $\bm{E}_a$ of type (B) and define $\bm{H}_a$ by
\begin{eqnarray}
\nabla\times\bm{E}_a=k_a\bm{H}_a.
\label{eq:ea}
\end{eqnarray}
Then one can check that $\bm{H}_a$ satisfies the equation (\ref{eq:bfun1}) and the boundary condition (\ref{eq:bfun2}), and hence belongs to type (B').
Conversely, starting from an eigenvector field $\bm{H}_b$ of type (B'), define $\bm{E}_b$ by
\begin{eqnarray}
\nabla\times\bm{H}_b=k_b\bm{E}_b.
\label{eq:ha}
\end{eqnarray}
Then the resulting $\bm{E}_b$ satisfies the equation (\ref{eq:efun1}) and the boundary condition (\ref{eq:efun2}), which characterize type (B).
These modes correspond to electromagnetic eigenmodes with nonzero frequencies ($\omega_n\coloneqq ck_n$).

\section{Mathematical formulas}
A minimal list of mathematical formulas is given below.
For details, see \cite{bib:sch95,bib:gro04,bib:eri17}.

\subsection{Riemannian metric}

A Riemannian metric on an $n$-dimensional manifold $M$ is a symmetric, positive definite bilinear form
\begin{equation}
g: T_pM\times T_pM\longrightarrow\R,\quad p\in M.
\end{equation}
In particular, for any nonzero $v\in T_pM$ we have $g(v,v)>0$.
Let $(x^1, \cdots, x^n)$ be a local coordinate system on $M$. The matrix
\begin{equation}
g_{ij}(p):=g\left(\frac{\partial}{\partial x^i}, \frac{\partial}{\partial x^j}\right),\quad
\frac{\partial}{\partial x^i},\frac{\partial}{\partial x^j}\in T_pM
\end{equation}
is symmetric and positive definite.
Let $G$ denote the determinant of $g_{ij}$, and let $g^{ij}$ denote the inverse matrix of $g_{ij}$; that is,
\begin{eqnarray}
&&G=\det(g_{ij}),\\
&&\sum_{j=1}^{n}g^{ij}(p)g_{jk}(p)=\delta^i_k.
\end{eqnarray}
Since $g$ is nondegenerate, it induces a linear map $\tilde{g}:  T_pM\longrightarrow T_p^{\ast}M$ defined by
\begin{equation}
\tilde{g}(u)(v)\coloneqq g(v,u),\quad u, v\in T_pM.
\end{equation}

\subsection{Hodge star operator}

Let $(y^1,\cdots,y^n)$ be a local coordinate system on $M$. We define the volume form $\mathrm{vol}^n$ on $M$ by
\begin{equation}
\mathrm{vol}^n\coloneqq \sqrt{|G|}\,dy^1\wedge\cdots\wedge dy^n.
\end{equation}
For a $p$-form $dy^{j_1}\wedge\cdots\wedge dy^{j_p}$ $(1\le j_1,\cdots,j_p\le n)$, we define its Hodge dual basis element $\vartheta_{j_1,\cdots, j_p}$ by
\begin{equation}
\vartheta_{j_1,\cdots, j_p}\coloneqq \mathrm{sgn}(\sigma)\sqrt{|G|}\,dy^{k_1}\wedge\cdots\wedge dy^{k_{n-p}},
\end{equation}
where $(k_1,\cdots,k_{n-p})$ is the sequence obtained from $(1,\cdots,n)$ by removing $(j_1,\cdots,j_p)$, and
$\sigma$ is the permutation that rearranges $(j_1,\cdots,j_p, k_1,\cdots,k_{n-p})$ into $(1,\cdots,n)$.
Using this, we define a linear map $\star: \Omega^p(M)\longrightarrow \Omega^{n-p}(M)$ as follows.
For a $p$-form
\begin{equation}
\omega=\sum_{1\le j_1,\cdots, j_p\le n}\omega_{ j_1,\cdots, j_p}\,dy^{j_1}\wedge\cdots\wedge dy^{j_p},
\end{equation}
we set
\begin{equation}
\star\omega\coloneqq
\sum_{1\le i_1,\cdots, i_p\le n}\sum_{1\le j_1,\cdots, j_p\le n}
\omega_{ i_1,\cdots, i_p}\,g^{i_1j_1}\cdots g^{i_pj_p}\,\vartheta_{j_1,\cdots, j_p}.
\end{equation}
This operator is called the Hodge star operator.
For $\omega\in\Omega^p(M)$, it satisfies
\begin{equation}
\star\star\omega=(-1)^{p(n-p)}\omega.
\end{equation}
In particular, $\star$ is an isomorphism.
We define the codifferential $\delta$ and the Laplacian $\Delta$ by
\begin{eqnarray}
&&\delta\coloneqq (-1)^{n(p+1)+1}\star d\star\\
&&\Delta\coloneqq d\delta+\delta d.
\end{eqnarray}
The operators $\star$ and $\Delta$ commute; that is,
\begin{equation}
\Delta\star=\star\Delta.
\end{equation}

\subsection{Integration of differential forms}

For a $(k-1)$-form $\omega$ and a $k$-chain $C$, we have
\begin{equation}
\int_C d\omega=\int_{\partial C} i^{\ast}\omega.
\end{equation}
Here $i^{\ast}$ is the pullback of the inclusion map $i:\partial C\rightarrow C$.
This is called Stokes' theorem.
For two $k$-forms $\omega$ and $\eta$, we define the inner product $(\omega,\eta)_M$ by
\begin{equation}
(\omega, \eta)_M\coloneqq\int_M \omega\wedge\star\eta.
\end{equation}
This inner product is symmetric and nondegenerate.
For a $(k-1)$-form $\alpha$ and a $k$-form $\beta$, we have
\begin{equation}
d(\alpha\wedge\star\beta)=d\alpha\wedge\star\beta+(-1)^k\alpha\wedge\star\delta\beta.
\end{equation}
Integrating this identity over $M$ and using properties of $\star$ together with Stokes' theorem, we obtain
\begin{equation}
(d\alpha, \beta)_M=(\alpha, \delta\beta)_M+\int_{\partial M} i^{\ast}(\alpha\wedge\star\beta).
\end{equation}

\subsection{Tangential and normal components of differential forms}
\label{subsec:boundary}
We explain how to describe boundary conditions for differential forms on the boundary $\partial M$ of a manifold $M$.
Let $i:\partial M\rightarrow M$ be the inclusion map.
For a $k$-form $\omega$, the condition
\begin{equation}
i^{\ast}\omega=0
\end{equation}
is called the Dirichlet boundary condition.
Also, the condition
\begin{equation}
i^{\ast}\star\omega=0
\end{equation}
is called the Neumann boundary condition.
Introduce local coordinates $(x^1,\cdots,x^n)$ near $\partial M$ so that $\partial M$ is given by $x^n=0$.
Then one can see that a form $\omega$ satisfying the Dirichlet boundary condition consists only of terms that contain $dx^n$, whereas a form satisfying the Neumann boundary condition contains no $dx^n$.

To study the behavior of $\omega\in\Omega^k(M)$ on $\partial M$, it is convenient to use the linear maps $\mathbf{t}$ and $\mathbf{n}$, which extract the tangential and normal components of a differential form, respectively.
To define them, first decompose a tangent vector $X\in T_pM$ at a point $p\in \partial M$ into the part orthogonal to $\partial M$ and the part tangent to $\partial M$:
\begin{equation}
X=X^{\perp}+X^{\parallel}.
\end{equation}
Using this decomposition, the tangential component $\mathbf{t}\eta$ of a $k$-form $\eta$ is defined by
\begin{equation}
\mathbf{t}\eta(X_1,\cdots,X_k)=\eta(X_1^{\parallel},\cdots,X_k^{\parallel}),
\quad X_1,\cdots,X_k\in TM|_{\partial M}.
\end{equation}
On the other hand, the linear map
$\mathbf{n}:~\Omega^k(M)|_{\partial M}\rightarrow \Omega^k(M)|_{\partial M}$
that extracts the normal component is defined by
\begin{equation}
\mathbf{n}\eta\coloneqq\eta|_{\partial M}-\mathbf{t}\eta.
\end{equation}
From the definition, we have
\begin{equation}
\mathbf{t}\cdot\mathbf{t}=\mathbf{t},\quad \mathbf{t}+\mathbf{n}=1.
\end{equation}
It is also easy to check that
\begin{equation}
\mathbf{t}\cdot\mathbf{n}=\mathbf{n}\cdot\mathbf{t}=0,\quad \mathbf{n}\cdot\mathbf{n}=\mathbf{n}.
\end{equation}

We list several useful properties of differential forms involving $\mathbf{t}$ and $\mathbf{n}$.
First, for a $k$-form $\eta$, the Hodge star operator $\star$ exchanges the normal and tangential components; that is,
\begin{equation}
\star(\mathbf{n}\eta)=\mathbf{t}(\star\eta),\quad \star(\mathbf{t}\eta)=\mathbf{n}(\star\eta).
\end{equation}
Moreover, the following identities hold:
\begin{equation}
d\,\mathbf{t}\eta=\mathbf{t}\,d\eta,\quad \delta\,\mathbf{n}\eta=\mathbf{n}\,\delta\eta.
\end{equation}
An important point is that, although $\mathbf{t}\eta$ and $i^{\ast}\eta$ are not the same object, we still have
\begin{equation}
\mathbf{t}\eta=0 \iff i^{\ast}\eta=0
\end{equation}
(\cite{bib:eri17}, Proposition 5.1).
Furthermore, using the properties above, we obtain
\begin{equation}
i^{\ast}\star \eta=0 \iff \mathbf{t}\star \eta=0 \iff \star \mathbf{n} \eta=0 \iff \mathbf{n} \eta=0.
\end{equation}
Therefore, the Dirichlet and Neumann boundary conditions can be expressed using $\mathbf{t}$ and $\mathbf{n}$.


\section{Maxwell equations with differential forms}

We rewrite the Maxwell equations and boundary conditions, originally written in the notation of three-dimensional vector calculus, using differential forms.
First, let the $1$-forms corresponding to the electric field $\bm{E}$ and the magnetic field $\bm{H}$ be
$\omega^1_E\coloneqq \tilde{g}(\bm{E})$ and $\omega^1_H\coloneqq \tilde{g}(\bm{H})$, respectively.
With this choice, it is natural to represent the electric flux density $\bm{D}$ and the magnetic flux density $\bm{B}$ by the $2$-forms
$\omega^2_D\coloneqq \epsilon_0\star \omega^1_E$ and $\omega^2_B\coloneqq \mu_0\star \omega^1_H$.
Then the Maxwell equations and boundary conditions can be written as follows:
\begin{eqnarray}
d\omega^1_E+\frac{\partial\omega^2_B}{\partial t}=0 \qquad &&{\mbox {in V}}\\
d\omega^1_H-\frac{\partial\omega^2_D}{\partial t} =0 \qquad &&{\mbox {in V}}\\
d\omega^2_D=0\qquad &&{\mbox {in V}}\\
d\omega^2_B=0\qquad &&{\mbox {in V}}\\
i^{\ast}\omega^1_E=0\qquad &&{\mbox {on S}}\\
i^{\ast}\star\omega^1_H=0\qquad &&{\mbox {on S}}.
\end{eqnarray}
Using properties of the Hodge star operator, this system can be written using only $\omega^1_E$ and $\omega^2_B$:
\begin{eqnarray}
d\omega^1_E+\frac{\partial\omega^2_B}{\partial t}=0 \qquad &&{\mbox {in V}}\\
\delta\omega^2_B-\frac{\partial\omega^1_E}{c^2\partial t}=0 \qquad &&{\mbox {in V}}\\
\delta\omega^1_E=0\qquad &&{\mbox {in V}}\\
d\omega^2_B=0\qquad &&{\mbox {in V}}\\
i^{\ast}\omega^1_E=0\qquad &&{\mbox {on S}}\\
i^{\ast}\omega^2_B=0\qquad &&{\mbox {on S}}.
\end{eqnarray}
Note that we now have only Dirichlet-type boundary conditions.
The wave equations for the electromagnetic field are then written as
\begin{eqnarray}
\label{eq:e1}
\delta d\omega^1_E-k^2\omega^1_E=0 \qquad &&{\mbox {in V}}\\
\label{eq:e2}
\delta\omega^1_E=0\qquad &&{\mbox {in V}}\\
i^{\ast}\omega^1_E=0\qquad &&{\mbox {on S}}\\
\label{eq:e4}
d\delta\omega^2_B-k^2\omega^2_B=0 \qquad &&{\mbox {in V}}\\
\label{eq:e5}
d\omega^2_B=0\qquad &&{\mbox {in V}}\\
i^{\ast}\omega^2_B=0\qquad &&{\mbox {on S}}.
\end{eqnarray}
When $k\neq 0$, (\ref{eq:e2}) and (\ref{eq:e5}) can be derived from (\ref{eq:e1}) and (\ref{eq:e4}), respectively.

\end{document}